\documentclass[journal,twoside]{IEEEtran}

\usepackage[english]{babel}

\usepackage{cite}
\usepackage{textcomp}

\usepackage[pdftex]{graphicx}
\graphicspath{{img/},{img/bio/}}

\usepackage{tikz}
\usepackage{lipsum}
\usepackage{color}

\usepackage[cmex10]{amsmath}
\usepackage{url}

\def\paperTitle{Tunable Eight-Element MIMO Antenna Based on the Antenna Cluster Concept}

\begin{document}
%
\title{\paperTitle}
%
%

\author{Jari-Matti~Hannula,
        Tapio~O.~Saarinen,
        Anu~Lehtovuori,
		Jari~Holopainen,
        and~Ville~Viikari
\thanks{The authors are with the Department of Electronics and Nanoengineering,
Aalto University School of Electrical Engineering, Espoo 02150, Finland
(e-mail: jari-matti.hannula@aalto.fi; tapio.o.saarinen@aalto.fi; anu.lehtovuori@aalto.fi; jari.holopainen@aalto.fi; ville.viikari@aalto.fi).}
}

\markboth{}%
{HANNULA \MakeLowercase{\textit{et al.}}: \MakeUppercase{Tunable Eight-element MIMO Antenna Based on the Antenna Cluster Concept}}

\maketitle

\begin{abstract}
Realizing capacity demands of future wireless communications
requires improved spectral efficiency in the sub-6-GHz frequency bands.
This paper proposes a novel eight-element
multiple-input multiple-output (MIMO) antenna that can be tuned from 1.7 to 6\,GHz.
The design is based on the antenna cluster concept,
where weighted feeding of multiple antenna elements is used to modify the operating frequency.
This paper extends the theory of the concept to account for multiple separate clusters,
thus enabling it to be used for MIMO.
The proposed antenna achieves over 60\% efficiency at frequencies above 3\,GHz,
and the system exceeds the ergodic capacity of the ideal 7$\times$7 MIMO in that band.
\end{abstract}


\section{Introduction}

Increasing expectations for wireless communications demand even better network solutions.
Mobile data traffic has grown 18-fold in the past five years and is expected to increase sevenfold during the next five years~\cite{cisco2017forecast}.
To answer to these demands, the fifth generation of mobile communications (5G) is being standardized and developed.
The 5G networks should provide 100--1000 times the capacity of the current Long-Term Evolution (LTE)-based fourth-generation (4G) networks~\cite{huawei20135g,nokianetworks2015looking,andrews2014what}. The fifth-generation communications standard, IMT-2020, is planned to be released by 2020.

To cover future demands, additional frequency bands have been allocated in 
an agreement signed in WRC15 \cite{wrc15spectrum}.
For example, 3.4--3.6\,GHz has been allocated globally for mobile communications.
3.6--3.8\,GHz is available in some regions, and there are ongoing discussions for allocating  the 3.8--4.2-GHz band for mobile use in Europe.
All in all, the allocated sub-6-GHz spectrum will be globally more diverse but also more fragmented,
demanding increased flexibility from the antennas.

The spectrum available below 6\,GHz will remain rather scarce even with further frequency allocations.
Therefore, the largest improvements in capacity must come from improving spectral efficiency, i.e., the information rate that can be transmitted over a given bandwidth.
The most significant improvements, especially from the antenna perspective, come from multiple-input multiple-output (MIMO) techniques.
The demand for increased capacity turns the research to higher order MIMO, which is already on the way.
$8 \times 8$ MIMO in the downlink and $4 \times 4$ in the uplink is included in the specifications of the LTE Advanced release 10.
Although there are already many publications on MIMO, the research focuses either on $2 \times 2$ or a high-order MIMO implemented in a narrow band.
Research on eight-element MIMO is mainly limited to the 3.5-GHz band~\cite{al-hadi2014eight,ban20164g5g,wong2017two} and above~\cite{li201812}.
In~\cite{li2016eight}, an eight-element MIMO antenna is investigated for the 2.5-GHz band.

In this work, we design an eight-element MIMO antenna in a handset form factor.
Previously published designs~\cite{al-hadi2014eight,ban20164g5g,wong2017two,li2016eight} are limited to a narrow band,
whereas in this work we design an antenna capable of operating at frequencies from 1.7 to 6\,GHz.
This operation across a wide bandwidth is enabled by the concept studied in~\cite{hannula2016concept,hannula2017frequency},
which involves the use of multiple closely spaced antenna elements to form an antenna cluster.
The operating frequency of the cluster can then be changed by weighting the signal fed to each element,
somewhat similar to beam-steering an antenna array.

This concept has previously only considered a single cluster.
In this work, we extend the theory to consider all antennas in the vicinity,
enabling us to minimize their effect and to use the concept for MIMO.
The work also illustrates the limitations of eight-element MIMO,
when applied to frequencies below 3\,GHz.
We show that although the antennas can be matched well,
the limitations of the chassis size will limit the overall performance.

This paper is organized as follows.
Section~\ref{sec:theory} explains the theory for both MIMO and the frequency reconfigurability.
The antenna design is shown in Section~\ref{sec:antenna}.
Finally, Section~\ref{sec:results} contains the results of the antenna performance,
where both the individual elements and the resulting MIMO performance are evaluated.

\section{Background of the Antenna Cluster Concept}
\label{sec:theory}

The antenna design presented in this paper uses the frequency reconfigurability concept in~\cite{hannula2016concept,hannula2017frequency},
which is also extended here.
This section briefly describes the concept to provide the necessary background for the proposed antenna design.
The fundamental principle of the method is the simultaneous use of multiple antennas to form an \emph{antenna cluster}.
Each antenna is fed the same data, but with specific amplitude and phase weighting.
By choosing the weights properly, both the mismatch and coupling can be reduced,
assuming the antennas are designed in a suitable manner.
The weights can then be adjusted to tune the antenna to another frequency.
The implementation of the weighting requires a suitable transceiver.
The transceiver implementation has been discussed in~\cite{hannula2018performance}.

The starting point for evaluating the performance of the antenna cluster is the formation of the radiation matrix\footnote{In~\cite{volmer2008eigen}, the radiation matrix is denoted by $\mathbf{H}$. However, so as not to confuse it with the notation for the MIMO channel $\mathbf{H}$ in this paper, we have adapted the convention of using $\mathbf{D}$ from~\cite{czawka2008new}, as in the dissipation matrix.} $\mathbf{D}$~\cite{volmer2008eigen,czawka2008new}.
If the antenna scattering parameters are known, the radiation matrix is
\begin{equation}
\mathbf{D} = \mathbf{I} - \mathbf{S}^\mathrm{H}\mathbf{S}.
\label{eq:dissmatrix}
\end{equation}
The radiation matrix characterizes the total performance of the antenna. If all the antenna excitations are defined with a vector $\mathbf{a}$, the efficiency $\eta$ of the antenna cluster is
\begin{equation}
\eta = \frac{\mathbf{a}^\mathrm{H} \mathbf{D} \mathbf{a}}{\mathbf{a}^\mathrm{H} \mathbf{a}}.
\end{equation}
To find the excitation that maximizes the efficiency of the cluster, the eigenvalues of $\mathbf{D}$ must be calculated.
The largest eigenvalue of $\mathbf{D}$ equals the efficiency of the antenna, when it is excited with the eigenvector corresponding to that eigenvalue, i.e.,
\begin{equation}
\eta_\mathrm{max} = \max \left\lbrace \mathrm{eig}\left( \mathbf{D} \right) \right\rbrace.
\label{eq:eigs}
\end{equation}
Because $\mathbf{D}$ changes as a function of frequency, the eigenvectors (and the equivalent feed coefficients) also vary with frequency.
Thus, by modifying the weights, the operation of the cluster can be tuned.
The efficiency results presented later in this paper show the maximum obtainable efficiency at each frequency point,
as opposed to the instantaneous bandwidth of the antenna.
The assumption is that the transceiver can provide the required weighting at each frequency point.

The results published on this topic so far assume that the cluster is the only antenna in the vicinity, and any power that is not reflected is radiated~\cite{hannula2016concept,hannula2017frequency}.
This is because the weighting coefficients have been calculated from the scattering parameters.
Scattering parameters cannot distinguish between radiated power and losses, as it only considers whether the power reflects back to the ports or if it is dissipated inside the network.
Thus, the radiation matrix $\mathbf{D}$ calculated from the scattering parameters can only be used to minimize the reflected power.
When the device has only one antenna cluster and low ohmic losses, this is not a problem.
With MIMO, however, the emphasis is also on the other antennas on the device.
If the effect of the other antennas is not considered, $\mathbf{D}$ gives only the matching efficiency,
so ignoring the effect of coupling, which is a significant source of loss in MIMO systems.
Additionally, the weighting coefficients might not be optimal for maximizing the radiated power.
This relates to the idea of uncoupled matching~\cite{jensen2010uncoupled}, in that one should not aim to minimize the reflections, but also consider the coupling to other antennas.

There are two ways to model the effect of the other antennas.
One option is to include every antenna on the device in the scattering matrix.
However, we do not want to feed all the antennas simultaneously, because that would combine all the antennas into one overly large cluster and prevent them from being used for MIMO.
Therefore the scattering matrix should only include the columns corresponding to the active antenna elements, i.e., the elements used to form the specific cluster, but also include the rows that describe how the active antennas couple to the other nearby antenna elements.

Alternatively, based on the relationship between the radiation patterns and the scattering matrix~\cite{stjernman2005relationship}, the radiation matrix can be calculated from the far-field radiation patterns.
The advantage of the far-field approach is that it considers all the losses in the system.
There is no need to explicitly include antennas outside the cluster in the radiation matrix,
and ohmic losses are also considered.
Using far-field patterns, the elements of the radiation matrix can be calculated from~\cite{stjernman2005relationship,volmer2008broadband}
\begin{equation}
D_{i,j} = \frac{1}{4\pi}\iint_{4\pi} F_i \cdot F_j^* \mathrm{d}\Omega
\label{eq:radmatrix}
\end{equation}
where $F_i$ is the radiation pattern of the $i$th antenna, normalized such that $D_{i,i}$ equals the total efficiency of that antenna.
The distinction between (\ref{eq:dissmatrix}) and (\ref{eq:radmatrix}) is that (\ref{eq:dissmatrix}) considers the matching efficiency of the cluster whereas (\ref{eq:radmatrix}) the total efficiency.
If the antenna structure is lossless, (\ref{eq:dissmatrix}) equals (\ref{eq:radmatrix}).

\section{Frequency Reconfigurable MIMO Antenna}
\label{sec:antenna}

\subsection{Antenna Design}

\begin{figure}[!t]
\centering
\includegraphics[width=\columnwidth]{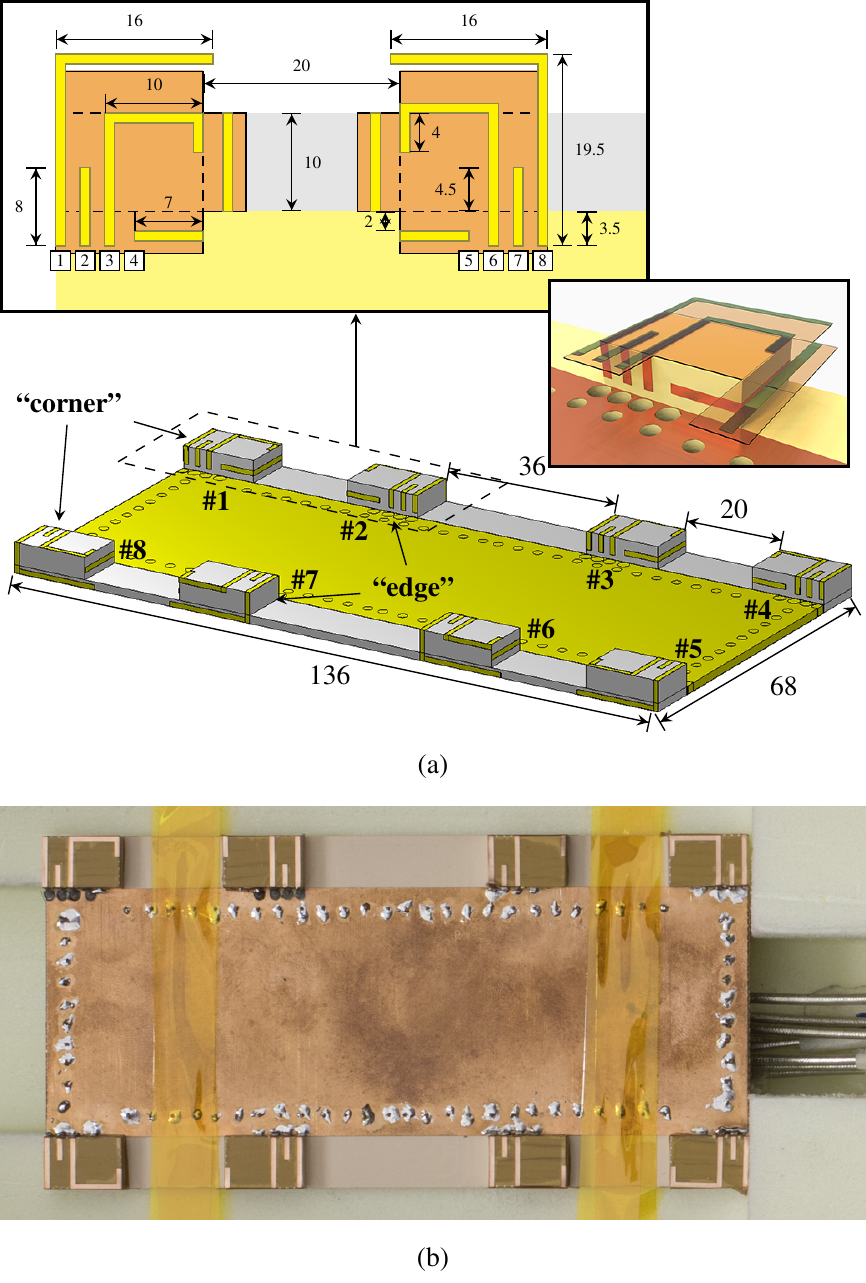}
\caption{The proposed MIMO antenna. (a) Dimensions of the antenna, showing both the overall structure and the cluster dimensions. All dimensions are in millimeters. The structure is symmetric along both axes. (b) The manufactured prototype.}
\label{fig:cst}
\end{figure}

The described theory is used to implement an $8 \times 8$ MIMO in a handset form factor.
The volume of the chassis is 136$\times$68$\times$6\,mm$^3$, corresponding roughly to that of a modern smartphone.
The design includes eight antennas, each with a volume of 15$\times$10$\times$6\,mm$^3$.
The antennas are placed along the long edges of the chassis, as shown in Fig.~\ref{fig:cst}.
Although the 10-millimeter side clearance is not compatible with current trends in smartphone design,
the proposed antenna can be used to demonstrate the feasibility of $8 \times 8$ MIMO at lower frequencies.

In this design, one antenna cluster consists of four antenna elements, each with a different resonant frequency.
To take advantage of the full volume allocated for each antenna, the antenna is designed on a substrate block with these dimensions.
The design of the individual antenna elements is done starting with the longest element that covers the lowest frequencies of the targeted operation bands.
The idea is to route the longest antenna alongside the edges of the block to obtain the required length in a reasonable manner without meandering.
A design principle is to keep the elements with approximately the same length as far away from each other as possible. 
Other elements are then added and spaced accordingly to obtain frequency reconfigurability across the whole band in accordance with the concept in~\cite{hannula2016concept}.
The dimensions of the antenna elements are given in Fig.~\ref{fig:cst}(a).
The design is based on the one presented in~\cite{hannula2017frequency}, which has been modified to better take advantage of the height of the antenna.

The ideal number of antenna elements depends on the targeted frequency range and available antenna volume. During the antenna design process, four proved to be the optimal number of antenna elements for covering the entire frequency range.
In our simulations, we were unable to cover the entire bandwidth with only three elements.
Adding a fifth element was not considered since the specifications were already achieved with four.
 
After designing a working antenna cluster, the block is duplicated to eight different locations.
A challenge with the block placement is that obtaining good matching at 1.7\,GHz requires that the block is placed as close to the short edge of the ground plane as possible to properly excite the dominant mode in the chassis~\cite{villanen2006coupling,martens2011inductive}.
However, because the corners are occupied by the four antenna blocks, the remaining four blocks must be placed further away from the corners.
The placement of the middle blocks is a design parameter and is modified to obtain the best possible efficiency.

\subsection{Prototype}

The prototype is manufactured on an RF plastic PREPERM 255 ($\varepsilon_\mathrm{r} = 2.55$, $\tan\delta = 0.0008$ at 1\,GHz) by Premix (\url{http://preperm.com}).
The substrate including the antenna blocks is milled to shape from a single piece of the plastic.
The ground plane is made from a thin sheet of laser-cut bronze which is attached to the substrate by routing and soldering wires through the via holes.
The measurement cables are routed through the back of the device to minimize their impact on the antennas.
For effective grounding and rigid connection of the measurement cables, the design includes a ground plane on both sides of the substrate.
The two ground planes are connected with several vias, as shown in Fig.~\ref{fig:cst}, to ensure that they are at the same potential.
The antenna elements are made using copper-coated Kapton polyimide, on which the antenna shape is created.
The polyimide has an adhesive back, so it is cut to shape, placed on the antenna block, and then bent to cover the sides of the block.

\section{Results}
\label{sec:results}

In this section, we show the performance results for the MIMO antenna prototype.
The results are divided into two categories: Intra-cluster results,
which show the performance of an individual cluster, and MIMO results,
where the operation of the entire antenna system is evaluated.

The analysis in this paper is performed without an actual reconfigurable transceiver.
The concept was experimentally verified in~\cite{hannula2017frequency}.
In~\cite{hannula2018performance}, the feasibility of an integrated transceiver implementation is discussed.
As of this moment, no suitable transceivers exist for this concept.
Thus the analysis is performed based on combining individually simulated and measured field data
using the theory presented in Section~\ref{sec:theory}.

\subsection{Intra-cluster Performance}

The design includes two types of antenna clusters depending on their location:
corner and edge clusters.
Fig.~\ref{fig:cst} shows the locations of the clusters.
Due to symmetry, only two clusters are required to characterize the entire antenna.
The remaining clusters have identical performance
with the only difference being in the orientation of the radiation pattern, which can be changed
by mirroring the pattern along the corresponding symmetry plane(s).

\begin{figure}[!t]
\centering
\includegraphics[scale=1]{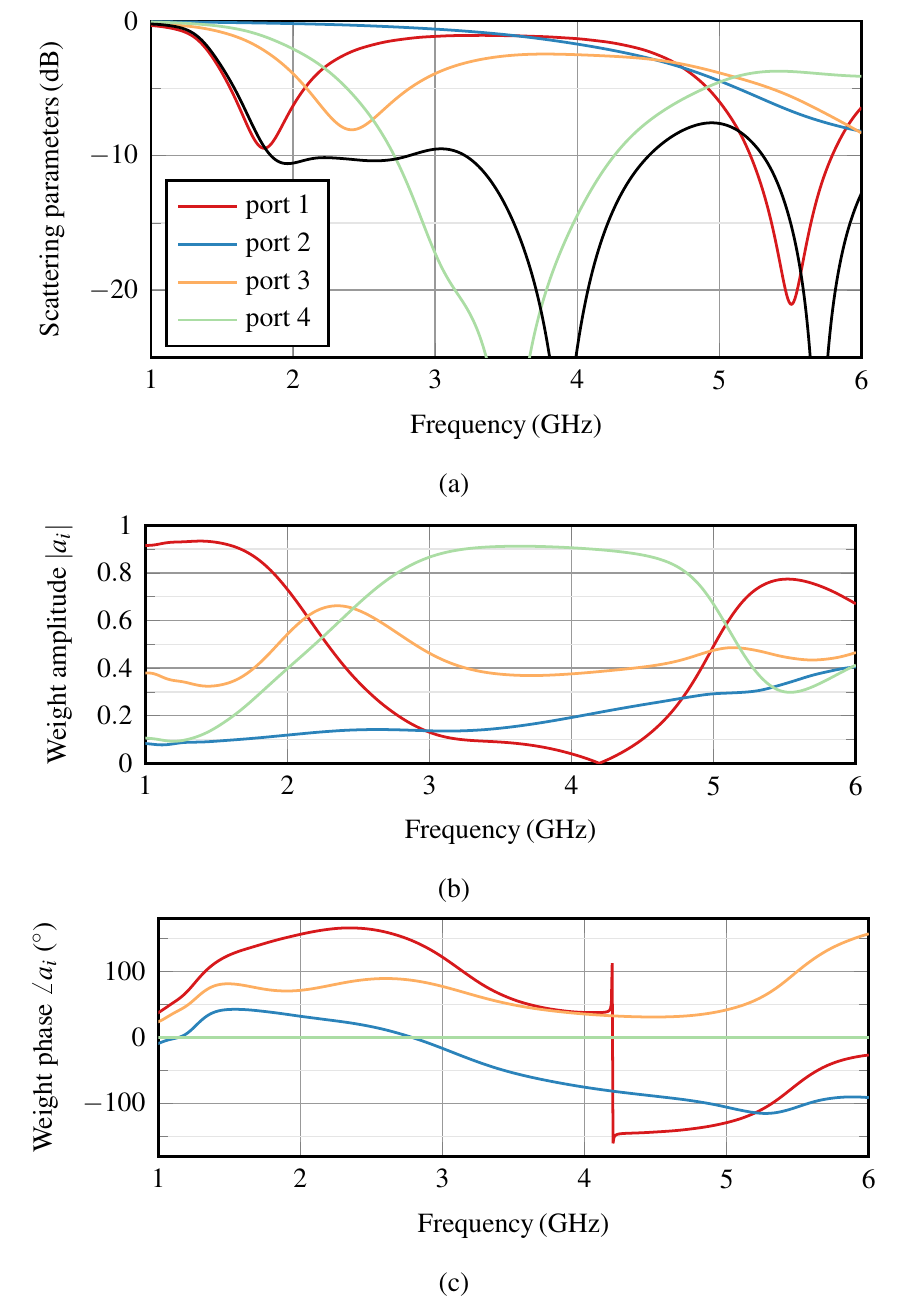}
\caption{Operation of the corner cluster: (a) the scattering parameters of the individual antennas and the resulting total active reflection coefficient including weights (black curve), and the corresponding weights (b) amplitude and (c) phase at each frequency point. Port numbering in all figures follows the convention shown in (a).}
\label{fig:indeff}
\end{figure}

Fig.~\ref{fig:indeff}(a) shows the scattering parameters for the corner cluster.
The behavior of the edge cluster is similar to that of the corner cluster
and is not depicted for brevity.
The colored curves depict the reflections $S_{ii}$ of each port.
The black curve depicts the minimum total active reflection coefficient (TARC), obtained when the
cluster is fed with the excitation shown in Fig.~\ref{fig:indeff}(b)--(c).
The excitation is to be generated with the transceiver, and changed according to the operating frequency,
thus making the cluster tunable.
Depending on the scattering parameters, sometimes the optimal excitation at a port can be zero.
This occurs around 4.1\,GHz, where feeding port one is not necessary since the combined coupling from the other ports cancels itself.

The results show that a very flat matching level is obtained from 1.8 to 3.2\,GHz,
and a TARC smaller than $-$10\,dB is reached over most of the frequency range.
The exception is around 5\,GHz,
which none of the elements cover very well.
Comparing the TARC to the reflection coefficients of ports 1 and 3 shows the benefit of the method.

\begin{figure}[!t]
\centering
\includegraphics[scale=1]{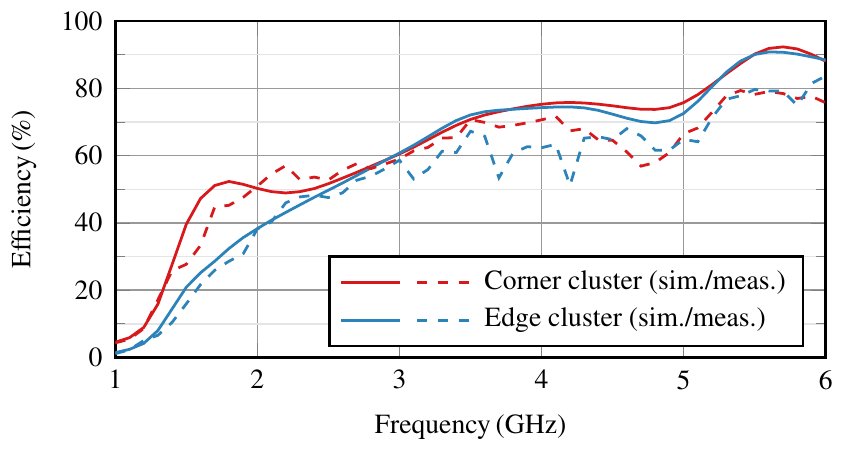}
\caption{Simulated and measured total efficiencies of the corner and edge clusters.}
\label{fig:eff}
\end{figure}

The far-field patterns of the antenna elements are measured using the \emph{MVG StarLab 6\,GHz} measurement system.
Each element is measured separately, with the remaining elements terminated with 50-ohm loads.
The radiation matrix is formed from the radiation patterns using~(\ref{eq:radmatrix}),
and optimal weights are computed from~(\ref{eq:eigs}).
Fig.~\ref{fig:eff} illustrates the total efficiencies of the corner and edge clusters when the weighted feeds are applied.
The simulated and measured efficiencies are in good agreement with each other.
Although both clusters resonate at the same frequency, the edge cluster operates with reduced efficiency.
The coupling to the lowest chassis mode is the strongest in the corner~\cite{villanen2006coupling,martens2011inductive},
which improves the performance of the corner cluster.
The edge element does not have this advantage, and thus reduced performance is obtained near the lower limit of the design bandwidth.

Using the scattering parameters and (\ref{eq:dissmatrix}) provides essentially identical results.
The intra-cluster coupling appears to be much more significant than the coupling to the other clusters.
As such, there is not much of a difference between calculating the weights from the far field or the scattering matrix,
at least in this structure.
The main benefit is the measurement process itself: the far-field approach requires eight measurements, whereas the scattering parameter approach requires 28 measurements with a two-port vector network analyzer.

The measurement results show that a total efficiency of 50\% is reached above 2.5\,GHz where the performance of the corner and edge clusters is basically identical.
This shows the importance of the chassis mode with regards to the overall performance in the traditional 1.7--2.7\,GHz band.
Efficient implementation of an $8 \times 8$ MIMO is therefore difficult in that band, because only half of the antennas can be placed in the corner.

Once the weighted far-field patterns have been generated, the cluster can be considered an ordinary single-element antenna for the purposes of evaluating the MIMO performance.

\subsection{MIMO Performance}

\begin{figure}[!t]
\includegraphics[width=\columnwidth]{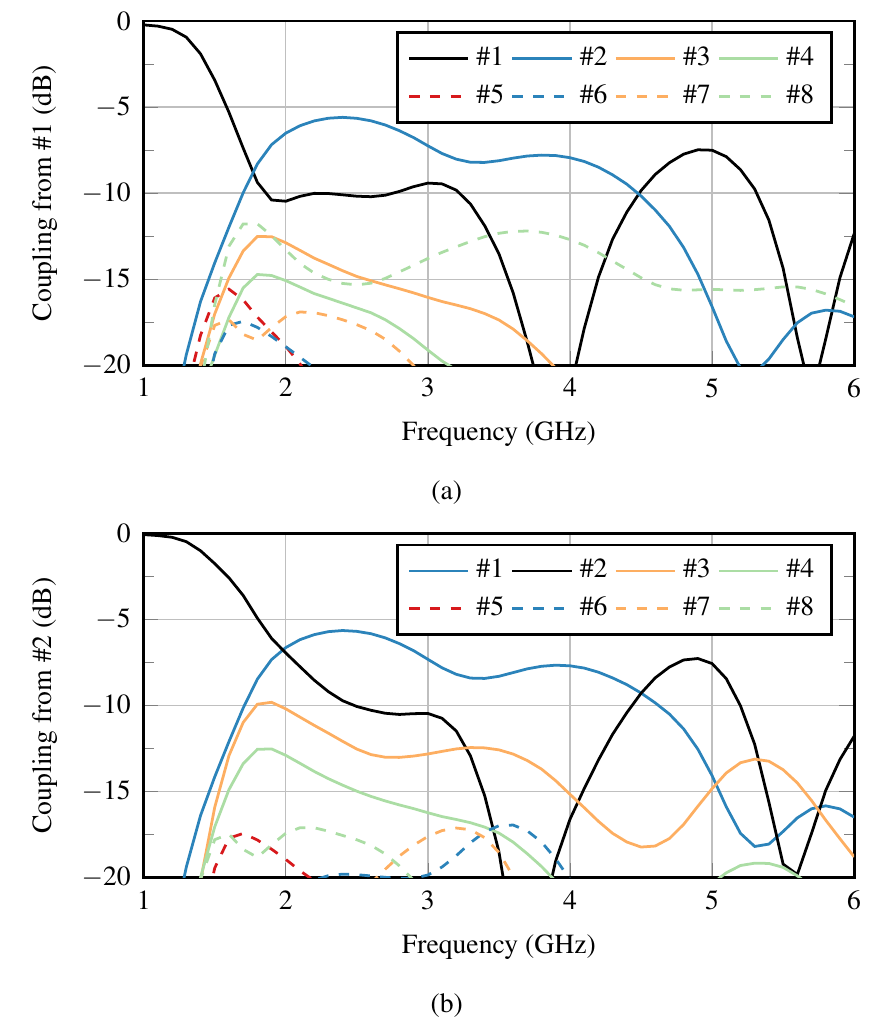}
\caption{Simulated coupling of power to different clusters from the (a) corner and (b) edge clusters. The black curves represent loss due to mismatch (coupling to itself).}
\label{fig:coupling}
\end{figure}

With the antenna clusters characterized, the MIMO performance of the antenna is analyzed using common metrics of MIMO described, e.g., in~\cite{paulraj2003introduction,tian2011multiplexing}.
Because only two clusters are measured, the patterns of the remaining clusters are generated by performing corresponding coordinate transformations on the measured patterns.

\begin{figure}[!t]
\includegraphics[width=\columnwidth]{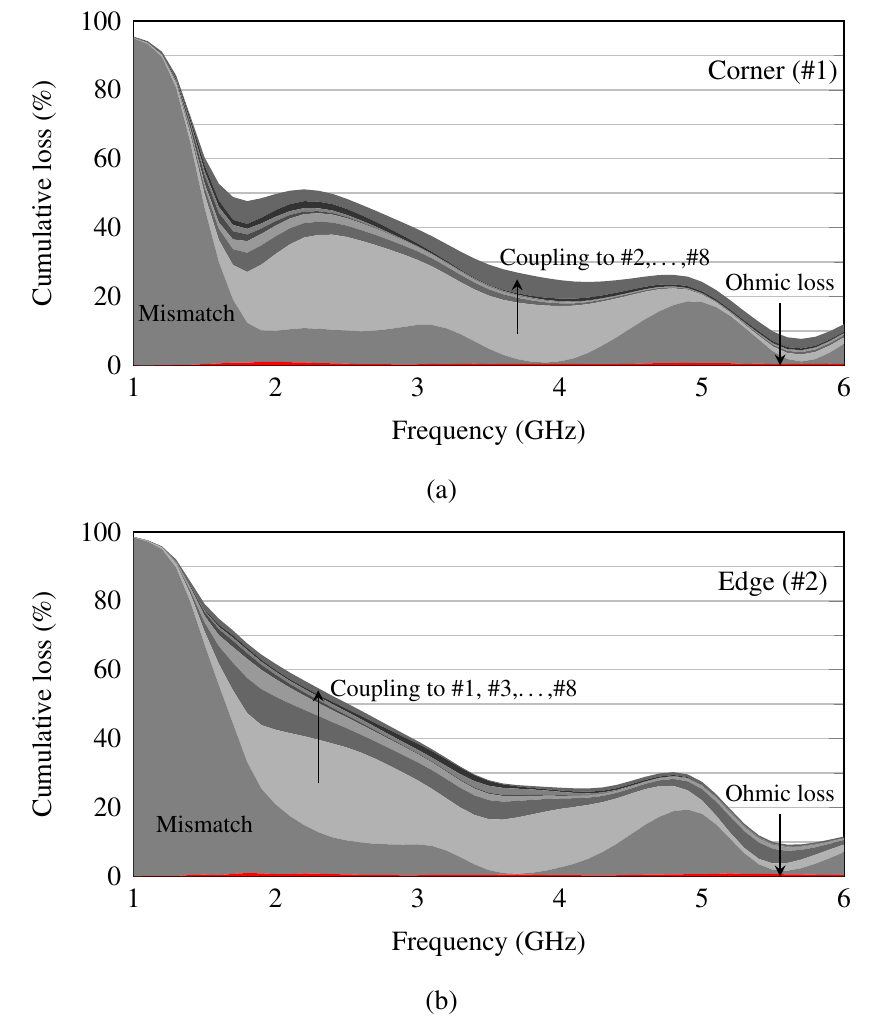}
\caption{Simulated sources of loss for a signal fed to the (a) corner and (b) edge clusters stacked on top of each other. Starting from the bottom, ohmic loss (red area, less than 1\,\%), mismatch loss, and coupling to other clusters in ascending order are shown.}
\label{fig:power}
\end{figure}

The first parameter that affects the MIMO performance is the efficiency of antennas,
shown already in Fig.~\ref{fig:eff}.
To analyze the loss of efficiency further,
Fig.~\ref{fig:coupling} shows the matching and coupling of both the corner and edge clusters.
The largest source of loss is quite evident from the result: the coupling between the adjacent corner and edge clusters, reaching even $-$6\,dB at some frequencies.
Additionally, the edge cluster has mismatch loss in the 1.7\,GHz band.
The coupling could be reduced by moving the edge clusters further away from the corners,
but this comes at the cost of reduced matching efficiency.
The chosen placement is a compromise between these two effects.

In the case of high-order MIMO,
the effect of the smaller coupling cannot be neglected either.
To demonstrate this behavior, Fig.~\ref{fig:power}(a) illustrates the different sources of loss for power fed to the corner cluster \#1.
This representation emphasizes the cumulative effect the coupling has.
Although the coupling to the individual non-adjacent clusters is well below $-$10\,dB,
the total coupling to them at 1.7\,GHz accounts for over 20\% of loss in efficiency.

At 1\,GHz, almost all power is lost due to the impedance mismatch.
As the frequency increases, the matching improves but the coupling grows as well.
Thus, from 1.7 to 2.5\,GHz the total loss remains quite constant.
The largest source of coupling also changes: as the coupling to clusters \#3--\#8 decreases,
the improved matching of cluster \#2 causes more power to couple to it from cluster \#1.
The adjacent cluster remains the largest source of loss (almost 30\,\% at 2.5\,GHz) all the way to 5\,GHz,
where its effect finally diminishes.

Fig.~\ref{fig:power}(b) depicts the same data for the edge cluster \#2.
The edge cluster shows a similar trend,
although with larger overall loss due to the impedance mismatch.
The loss curves of both clusters converge at around 3\,GHz,
as is evident also from the efficiency curves depicted in Fig.~\ref{fig:eff}.

\begin{figure}[!t]
\includegraphics[width=\columnwidth]{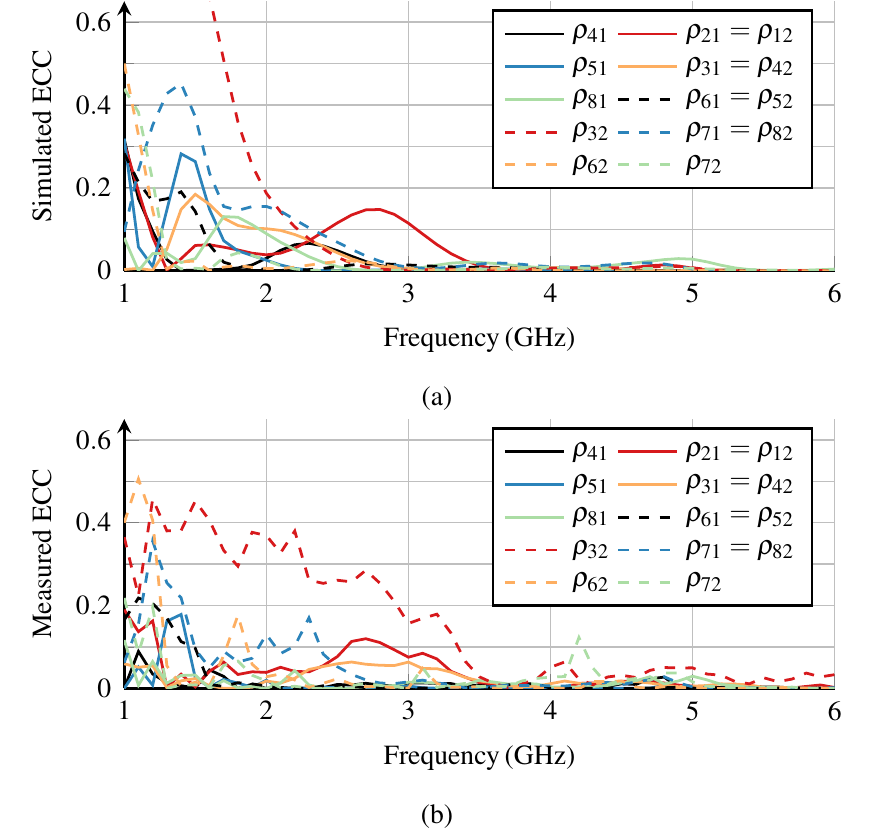}
\caption{Envelope correlation coefficients calculated from (a) simulated and (b) measured far-field patterns.}
\label{fig:ecc}
\end{figure}

In addition to efficiency, the correlation between the antennas has an effect.
Fig.~\ref{fig:ecc} depicts the Envelope Correlation Coefficients (ECC) calculated from the simulated and measured far-field data.
The ECC is under 0.2 in the operating band, with the exception of the correlation between the adjacent edge elements ($\rho_{32})$.
Nevertheless, in the operating band the ECC is below 0.5, which is considered adequate for MIMO operation~\cite{tian2011multiplexing}.

\begin{figure}[!t]
\includegraphics{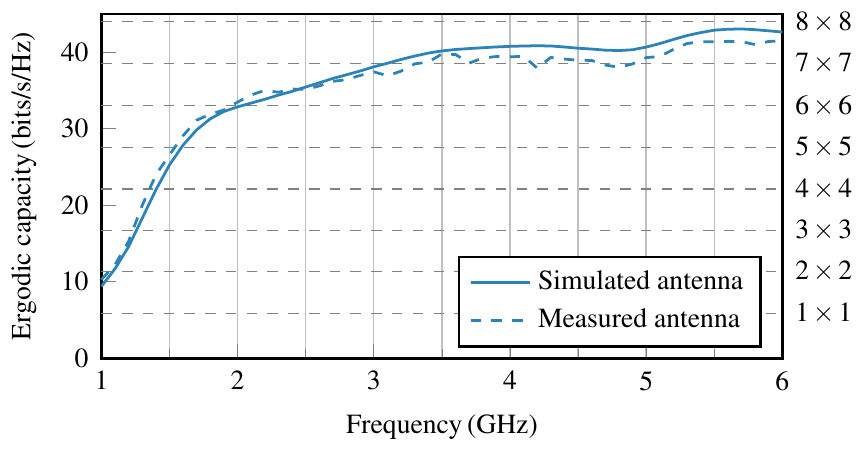}
\caption{Calculated ergodic capacity for both the simulated and measured antenna parameters when the SNR = 20\,dB. The ideal capacity for different $M \times M$ MIMO channels is shown for comparison.}
\label{fig:capacity}
\end{figure}

Fig.~\ref{fig:capacity} shows the ergodic capacity, calculated from using $10^4$ random samples at each frequency point.
A Rayleigh fading channel is assumed for the propagation environment.
In the calculations we assume the SNR to be 20\,dB and the number of base station antennas to be equal to the number of antennas in the handset.
Above 3\,GHz, the antenna enables a capacity larger than that of the ideal $7 \times 7$ MIMO, reaching at best 97\% of the ideal capacity of the $8 \times 8$ MIMO.
Below 3\,GHz the capacity is limited because of reduced efficiency.

The theoretical limitations of MIMO capacity as a function of antenna size have been studied in~\cite{ehrenborg2018fundamental}.
The results of this work also indicate a limitation on capacity due to the limited chassis size.
Additionally, the behavior of the capacity curve is similar to that presented in~\cite{holopainen2017study},
where the MIMO capacity is studied by conjugate matching of the antennas at point frequencies.

\section{Conclusion}

This paper presented an eight-element MIMO antenna whose operating frequency is tunable over a wide bandwidth, from 1.7 to 6\,GHz.
The presented antenna is based on the antenna cluster concept, 
which was extended to MIMO antennas in this work.
Above 3\,GHz, over 60\% efficiency is obtained for each antenna,
resulting in an ergodic capacity exceeding that of the ideal 7$\times$7 MIMO.
At the highest frequencies, the achieved ergodic capacity is 97\% of the theoretical maximum.
This demonstrates that the desired improvement in spectral efficiency can be obtained with high-order MIMO
at these frequencies.
However, the results show that although the antennas can be matched well, obtaining good efficiency with an eight-element MIMO in the 1.7--2.7\,GHz band is challenging due to coupling.
This suggests that eight-element MIMO in this band is feasible at some level, although the loss in efficiency raises the question of whether
the limited improvements are worth the added complexity
or should MIMO in this band be limited to 4$\times$4.

\section*{Acknowledgment}

The authors would like to thank senior laboratory technician E.~Kahra for designing the antenna manufacturing approach
and building the antenna supports for the far-field measurements.

\IEEEtriggeratref{5}


\bibliographystyle{IEEEtran}
\bibliography{IEEEabrv,references}

\end{document}